\documentclass[a4paper, 12pt]{article}

\usepackage[latin1]{inputenc}
\usepackage[english]{babel}

\usepackage{graphicx}
\usepackage{amsmath}
\usepackage{latexsym}
\usepackage{amssymb} 
\usepackage{epsfig}
\usepackage{cancel}
\usepackage{exscale}

\begin{document}

\begin{center}
{\LARGE \bf A new type of critical behaviour in random matrix models\\  
\vspace{1cm}}

{ \large R. Flume and  A. Klitz}
\\ 
\vspace{0.5cm}
Physikalisches Institut, Universit\"at Bonn,\\
Nu{\ss}allee 12, 53115 Bonn, Germany\\ 
\vspace{0.5cm}

\begin{abstract}
In view of the remarkable progress in the analysis of random matrix
models during the last years we report on recent work of Eynard on the
generation of a new cut in finite distance from the cuts given in the
initial theory. This 'birth of a cut' leads to a third order phase 
transition. 
\end{abstract}

\end{center}

Given the long history of random matrix models with its almost innumerable
applications in many diverse fields of the sciences (for a review see \cite{one})
it is fascinating to observe that the subject provides again and again innovative surprises---both with respect to foundational mathematical aspects
as well as with respect to new applications. On the foundational level a
dichotomy of views has been adapted since the seminal work \cite{two}: one has
to distinguish between convergent and formal matrix models. As regards the former, one specifies the contours in the matrix integrals such that those are
indeed convergent. The evaluation of the integrals via saddle point techniques
includes then the task of finding the extrema with respect to the variation of
all relevant parameters. The problem simplifies somewhat if one considers
matrices of large size---what is commonly called the large $ N $ limit. The
spectra of the matrices in the large $ N $ limit are expected
to concentrate on a compact support. If
there are several disconnected components, one has to specify their relative
spectral weights, such that the chosen saddle point configuration is extremal with
respect to the latter. A basic insight in reference \cite{two} is that the
extremalization in the spectral weights generically gives rise to oscillatory
terms in $ 1 / N $. The net result is that the $ 1 / N $ expansion
of the matrix model breaks down. A systematic incorporation of
oscillatory terms to all orders in $ 1 / N $ has been proposed in
\cite{four}.
\\ \\
Formal matrix  models in contrast are $
1 / N $ expandable, as one is working without an extremalization in the
spectral weights, assuming instead ad hoc values for the latter. Formal matrix
models are mainly used as 'counting devices', that is, they yield formal
series representative for e.g. the number of graphs in topological expansions
or the enumerative description of some algebraic geometry invariants
etc. There has been spectacular progress in the field recently. It has been
shown in \cite{five} that the topological ($ 1 / N $) expansion for the
matrix model free energy can be generalized in the following sense: The
topological expansion in the matrix model case is neatly related to an ensuing
spectral curve, which in simple cases is a hyperelliptic Riemann surface. All
terms of the expansion can be expressed through the hyperelliptic differential
and the so-called Bergmann kernel $ B(p,q) \; dp\,dq $, a bidifferential on
the hyperelliptic surface with a double pole at the coincidence point $ p=q
$. Eynard and Orantin \cite{five} show that the functional relations for the free
energy can be extended to more general Riemann surfaces---no longer related to
matrix models---giving rise to meaningful symplectic invariants of these
surfaces and possibly solving some long standing mathematical problems
\cite{six}, \cite{seven}. In the article \cite{eigth} on 'the birth of a cut' Eynard is
dealing with a problem out of the realm of convergent matrix models. The
problem that he considers is depicted in figure 1.

\begin{figure}[h!] 
\begin{center}
\hspace{1.1cm}
\epsfig{figure=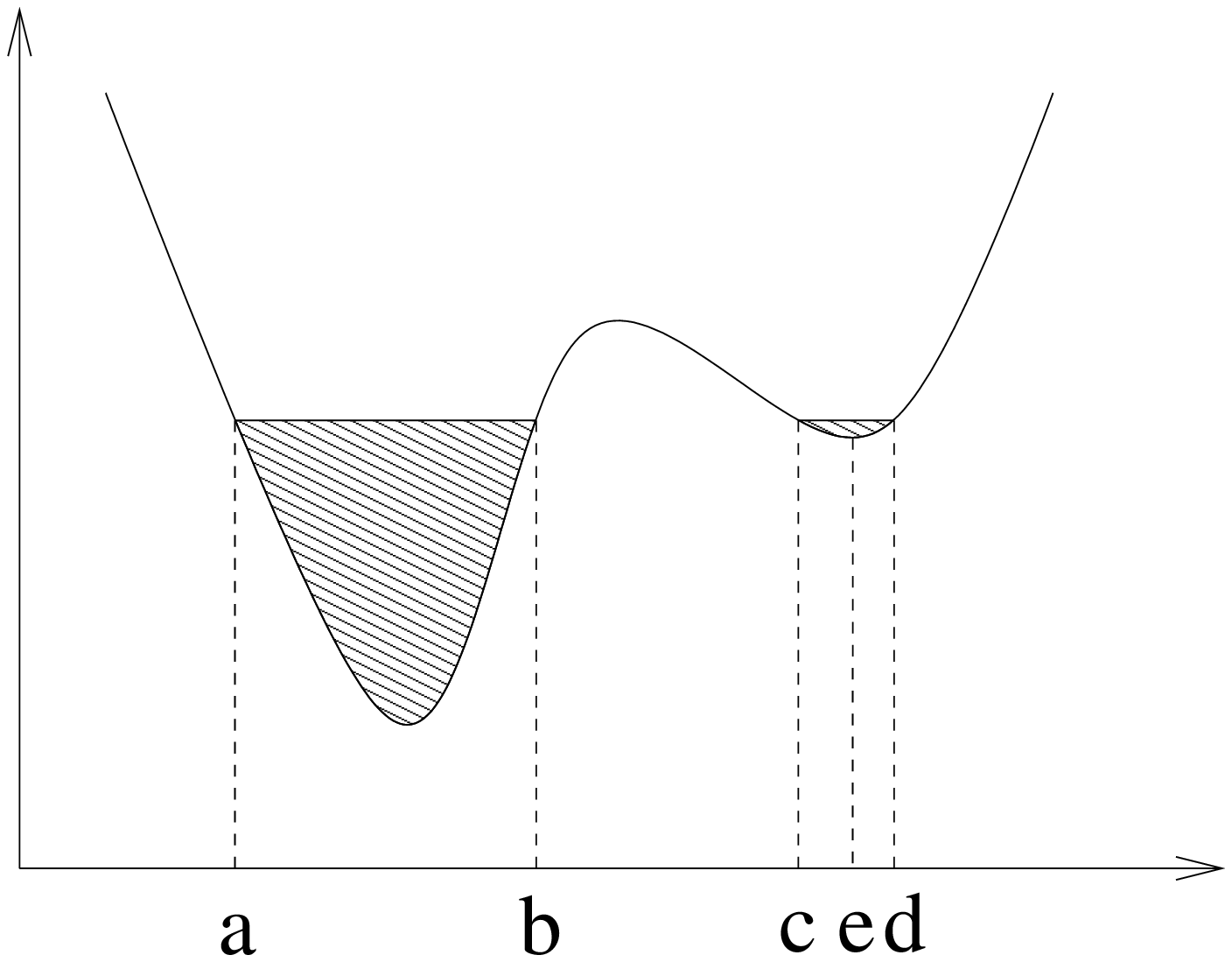, width=5cm} 
\caption{Birth of a cut} 
\end{center}
\end{figure}

The minima of a given matrix model potential (two in the case depicted) are
tuned such that (below a certain critical temperature) the local minima,
separate from the absolute minimum, are above the Fermi level of the
eigenstates around the absolute minimum. We are here in the situation of a
'one-cut' model, that is the eigenvalues of the matrix model are distributed
on a single connected interval---the interval $ [ a , b ] $ in figure 1. One
may now change one parameter of the model, call it a temperature, such that the
height of a second minimum passes through the Fermi level above the absolute
minimum. It is to be expected that above the 'critical temperature' (if the
second minimum is below the Fermi level attached to the first minimum) a
second cut around point $ e $ in figure 1 will become occupied by eigenvalues---a
new cut is born. The particular feature of the setup consists in the fact that a new cut is generated in finite distance from a
pre-existing cut: this situation differs from those considered in the
literature so far, where critical behaviour is related to merging and splitting of cuts
and is brought into correspondence with non-linear integrable hierachies and
rational conformal field theory \footnote{The author of reference
  \cite{Akemann} pointed to the fact that he made an observation already some
  time ago relevant in this context: A logarithmic scaling behaviour in the
  case of a vanishing cut.}. \\ \\
In a first, semiclassical, approach the transition from the one-cut case to
the two-cut case is modelled on the resolvent of the respective matrix
model. The resolvent is found to be related through the 'loop equation' to the
matrix model potential (for a review of the loop equation technique, see \cite{nine}). Below the critical temperature one is led to take for the
resolvent $ W $ the ansatz 
\begin{displaymath}
 W(x,T) \big|_{ T < T_c } = \frac{ 1 }{ 2 }
 \left( V^\prime (x) - M_- (x,T) \sqrt{ (x-a)(x-b) } \right) 
\end{displaymath}
 with $ a( T ) $ and $ b( T ) $, $ a( T ) < b ( T ) $, designating the endpoints of the cut below the critical
temperature, and a polynomial
\begin{displaymath}
 M_- (x,T_c) = (x-e)^{2\nu-1} Q(x),
\end{displaymath}
$ \nu $ denoting an integer and $ Q (x) $ being a real polynomial (with very
special properties which we will not go into). The suggestive ansatz for the
resolvent above the critical temperature, in the two-cut region, is 
\begin{displaymath}
 W(x,T) \big|_{ T > T_c } = \frac{ 1 }{ 2 }
\left( V^\prime (x) - M_+ (x,T) \sqrt{ (x-a)(x-b)(x-c)(x-d) } \right), 
\end{displaymath}
\begin{displaymath}
c(T_c) = d(T_c) = e \mbox{ \ and \ }  M_+ (x,T_c) = (x-e)^{2\nu-2} Q(x) .
\end{displaymath}
Starting from this ansatz one finds a smooth transition of the free energy and
its first and second derivatives with respect to temperature, but a jump in the
third derivative, that is, in usual nomenclature, a third-order transition.
\\ \\ A more detailed examination of the critical behaviour is performed after
integration of the eigenvalues in the low temperature cut concentrating on the
relations in the newly born cut. The interactions between the eigenvalues in the old and the new cut are taken care of in mean
field approximation. It is found that the relevant quantities going along with
a description in terms of orthogonal polynomials can be expressed in leading
order through the integer $ \nu $ introduced above and through the distance, $
b-e $, of the new from the old cut.\\
In \cite{ten}-\cite{twelve} three groups almost
simultaneously used the Riemann-Hilbert method to rigorously prove the results obtained in \cite{eigth}.
\\ \\
Whether a conformal field theory can be assigned to this critical behaviour is
unclear at the present time.


\begin{thebibliography}{99}
\bibitem{one} Guhr T, M\"uller-Groeling A and Weidenm\"uller H A, \it Random Matrix
  Theories in Quantum Physics: Common Concepts, \rm 1998 Phys.Rept. 299 189 [arXiv:cond-mat/9707301]
\bibitem{two} Bonnet G, David F and Eynard B, \it Breakdown of universality in
  multi-cut matrix models, 2000 \rm J.Phys. A: Math. Gen. 33 6739 [arXiv:cond-mat/0003324]
\bibitem{four} Eynard B, \it Large $ N $ expansion of convergent matrix integals,
  holomorphic anomalies, and background independence, 2008 \rm arXiv:0802.1788 [math-ph]
\bibitem{five} Eynard B and Orantin N, \it Invariants of algebraic curves and
  topological expansion, 2007 \rm arXiv:math-ph/0702045
\bibitem{six} Eynard B, \it Recursion between Mumford volumes of moduli
  spaces, 2007 \rm
  arXiv:0706.4403 [math.AG]
\bibitem{seven} Bouchard V and Mari$\tilde{\rm n}$o M, \it Hurwitz numbers, matrix models and
  enumerative geometry, 2007 \rm arXiv:0709.1458 [math.AG]
\bibitem{eigth} Eynard B, \it Universal distribution of random matrix eigenvalues
  near the 'birth of a cut' transition, 2006 \rm J. Stat. Mech. P07005 [arXiv:math-ph/0605064] 
\bibitem{nine} Di Francesco P, Ginsparg P and Zinn-Justin J, \it 2D gravity and
  random matrices, 1995 \rm Phys. Rept. 254 1 [arXiv:hep-th/9306153]
\bibitem{ten} Claeys T, \it Birth of a cut in Unitary Random Matrix
  Ensembles, 2008 \rm Int. Math. Res. Notices 2008 rnm166 \rm arXiv:0711.2609 [math-ph]
\bibitem{eleven} Mo M Y, \it The Riemann-Hilbert approach to double scaling
  limit of random matrix eigenvalues near the 'birth of a cut' transition,
  2007 \rm arXiv:0711.3208 [math-ph] 
\bibitem{twelve} Bertola M and Lee S Y, \it First Colonization of a Spectral
  Outpost in Random Matrix Theory, 2007 \rm arXiv:0711.3625 [math-ph]
\bibitem{Akemann} Akemann G, \it Universal correlators for multi-arc complex
  matrix models, 1997 \rm Nucl. Phys. B507 475 [arXiv:hep-th/9702005] 
\end{thebibliography}
\end{document}